\documentstyle[psfig]{mn2e}

\title[PHL~1092 observed with {\em XMM-Newton}]
{An intense soft-excess and evidence for light bending in the luminous narrow-line quasar PHL~1092
 }

\author[L. C. Gallo et al.]
{L. C. Gallo,$^1$ Th. Boller,$^1$ W. N. Brandt,$^2$ A. C. Fabian,$^3$ and D. Grupe$^4$ \\
$^1$ Max-Planck-Institut f\"ur extraterrestrische Physik, Postfach 1312, 85741 Garching, Germany \\
$^2$ Department of Astronomy and Astrophysics, The Pennsylvania State University, 525 Davey Lab, University Park, PA 16802, USA\\
$^3$ Institute of  Astronomy, Madingley Road, Cambridge CB3 0HA\\
$^4$ Department of Astronomy, The Ohio State University, 140 West 18th Avenue, Columbus, OH 43210, USA \\
}

\date{Accepted. Received. }

\begin{document}
\label{firstpage}
\maketitle

\begin{abstract}
The narrow-line quasar PHL~1092 was observed by {\em XMM-Newton} at two
epochs separated by nearly thirty months.  Timing analyses confirm the
extreme variability observed during previous X-ray missions.
A measurement of the radiative efficiency is 
in excess of what is expected from a Schwarzschild black hole.
In addition to the rapid X-ray variability, the short UV light
curves ($<$ 4 hours) obtained with the Optical Monitor may also show 
fluctuations, albeit at much lower amplitude than the X-rays.
In general, the extreme variability is impressive considering that the 
broad-band (0.4--10~keV rest-frame) luminosity of
the source is $\sim$ 10$^{45}$ erg s$^{-1}$.
During at least one of the observations, the X-ray and UV light curves show
common trends, although given the short duration of the OM
observations, and low significance of the UV light curves it is
difficult to comment on the importance of this possible correlation.
Interestingly, the high-energy
photons ($>$ 2~keV) do not appear highly variable.
The X-ray spectrum resembles that of many narrow-line Seyfert 1 type galaxies: an intense soft-excess 
modelled
with a multi-colour disc blackbody, a power-law component,
and an absorption line at $\sim$ 1.4~keV.
The $\sim$ 1.4~keV feature is curious given that it was not
detected in previous observations, and its presence could be related
to the strength of the soft-excess.
Of further interest is 
curvature in the spectrum above $\sim$ 2~keV which can be described by a
strong reflection component.
The strong reflection component, lack of high-energy temporal 
variability,
and extreme radiative efficiency measurements can
be understood if we consider gravitational light bending effects close to a
maximally rotating black hole.

\end{abstract}

\begin{keywords}
galaxies: active, AGN -- galaxies: individual: PHL~1092 --
X-rays: galaxies

\end{keywords}

\section{Introduction}
The narrow-line quasar PHL~1092 ($z = 0.396$) was observed
by {\em XMM-Newton} as part of the Guaranteed Time Program
to study Narrow-Line Seyfert~1 type objects (NLS1).
The importance of PHL~1092 was realised by Bergeron \& Kunth (1980)
when it appeared unique among a sample of quasars due to its 
outstanding Fe~{\tt II} 
emission.
During the $ROSAT$ era PHL~1092 was
recognised as a high-luminosity analogue of the NLS1
phenomenon (Forster \& Halpern 1996; Lawrence et al. 1997).
Remarkable variability
was observed during an 18-day \em ROSAT\em~HRI monitoring campaign
(Brandt et al. 1999; hereafter BBFR), including a number of high-amplitude
flaring events.  The most extreme variability
was an increase in the rest-frame count rate by nearly a factor of four in
less than 3580~s, corresponding to  a radiative efficiency (Fabian 1979; see also
Section 4.1)
of $\eta$ $>$ 0.63.  Such extreme variability had never before
been observed in such a high-luminosity radio-quiet quasar.

The objectives of this study are to determine if PHL~1092 consistently
exhibits such extreme behavior, and to constrain better the
physical nature of its X-ray spectral energy distribution.

\section{Observation and data reduction}
PHL~1092 was observed with {\em XMM-Newton}
(Jansen et al. 2001) on two separate occasions.
The first observation occurred on 2000 July 31 during revolution 0118 and
lasted for $\sim$ 32~ks.  During this time the EPIC pn (Str\"uder et al. 
2001)
and MOS (MOS1 and MOS2; Turner et al. 2001) cameras, as well
as the Optical Monitor (OM; Mason et al. 2001) and the Reflection Grating
Spectrometers (RGS1 and RGS2; den Herder et al. 2001) collected data.
From this first observation EPIC, OM, and RGS event files 
were created and supplied by the {\em XMM-Newton} Science Operations Centre.
On analysing the EPIC data it was found that problems with the energy
calibration existed.
As such, the EPIC spectra could not be exploited due
to calibration uncertainties within the small energy bins utilised for spectral
analyses.  Fortunately, broad-band light curves are not as sensitive to 
calibration uncertainties as spectra; hence it was possible to construct a 
pn light curve 
from these data.  
RGS and OM files were not affected.
Since the Observation Data Files (ODFs) could not be recovered for this
observation
PHL~1092 was re-observed on 2003 January 18 during AO2 (revolution
0570) for $\sim$ 28.5~ks.  
During this second observation all instruments
were functioning, and the ODFs were successfully produced.
At both epochs, the EPIC instruments used the medium filter and were 
operated
in full-frame mode.

The ODFs
were processed to produce calibrated event lists using the {\em
XMM-Newton} Science Analysis System ({\tt SAS v5.4.1}). Unwanted hot,
dead, or flickering pixels were removed as were events due to
electronic noise.  Event energies were corrected for charge-transfer
losses.  EPIC response matrices were generated using the {\tt SAS} tasks
{\tt ARFGEN} and {\tt RMFGEN}.
Light curves were extracted from these event lists to
search for periods of high background flaring.
High-energy background flaring was found to be extensive resulting in a loss
of $\sim$ 25\% of the data.  The total amount of good exposure
time selected was 18795 s.  
The source plus background photons were extracted from a
circular region with a radius of 35$^{\prime\prime}$, and the background was
selected from an off-source region with a radius of 50$^{\prime\prime}$, 
and appropriately scaled to the source
region.  Single and double events were selected for the pn
detector, and single-quadruple events were selected for the MOS.
The total number of counts collected
by the pn instrument in the 0.3--10~keV range was 12399.  In Table~\ref{cnts}
we give a distribution of the counts with respect to energy.
\begin{table}
\begin{center}
\begin{tabular}{lcccc}
 & 0.3 $\le E \le$ 2 & 2 $<$ $E$ $<$ 4 & 4 $\le E$ $<$ 7.2  \\
\hline
Source + & 11888 & 247 & 203 \\ 
background & & & & \\
Background  & 200 & 40 & 78 \\
\hline
\end{tabular}
\caption{Source plus background and background pn counts collected in
various energy bins.
Energies are given in units of keV.
The background counts have been scaled to the source cell size.
}
\label{cnts}
\end{center}
\end{table}
The {\em XMM-Newton} observation provides a vast improvement in spectral
quality over the 72.2~ks $ASCA$ exposure (199.6~ks duration) in
which $\approx$ 2900 counts were collected (Leighly 1999a).  In addition,
{\em XMM-Newton} is sensitive at lower energies than $ASCA$ was,
and this is critical in analysing this very soft source.
Although $ROSAT$ was capable of observing even lower energies than
{\em XMM-Newton}, only 2235 counts were collected in the 0.1--2~keV range
during the PSPC observation
(Forster \& Halpern 1996; Lawrence et al. 1997).

The RGS event lists were also created from the ODFs following standard {\tt 
SAS}
procedures.  However, it was determined that the RGS data from both epochs
were background dominated and would not be useful for this analysis.

The Optical Monitor collected data through the UVW2 filter
(1800--2250 \AA) for about the first 12~ks of each observation, and then it 
was switched
to UV grism mode for the remaining time.
In total, five photometric images were taken at each epoch.
The exposure times during the 2000 June observation were 2300~s, and during
the 2003 January observation they were 2640~s.

\section{X-ray spectral analysis}

\begin{figure}
\psfig{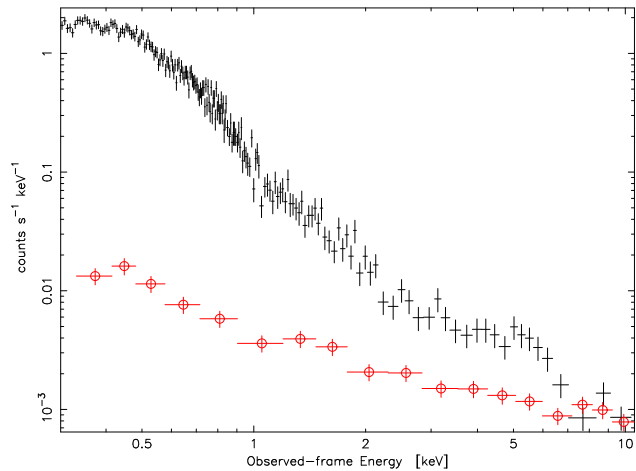}
\caption{The pn source and background spectra.  The upper curve
(black crosses) is the source plus background spectrum.
The lower curve (red, open circles) is the background spectrum.  The data above
7.2~keV are ignored as the source spectrum becomes background dominated.
}
\label{srb}
\end{figure}

Each of the EPIC spectra was compared to the
respective background spectrum to determine in which energy range the source
could be reasonably detected above the background.
The data were determined to be source dominated up to energies of
$\sim$7.2~keV
($\sim$10 keV in the rest-frame; Figure~\ref{srb}).
In addition, the MOS data were ignored below 0.7~keV due to the
uncertainties
in the low-energy redistribution characteristics of the cameras (Kirsch
2003).
When fitted with a power-law, the MOS1 data above 2.5~keV displayed a
steeper slope compared to the equivalent pn and MOS2 slopes ($\Delta\Gamma
\sim$ 0.20--0.25).
This inconsistency in the MOS1 data was previously realised in
observations of 3C~273 by Molendi \& Sembay (2003).
Molendi \& Sembay noted a difference in the MOS1 photon index compared to the
other EPIC photon
indices of $\Delta\Gamma$ $\sim$ 0.1.
We note that since the PHL~1092 high-energy spectra are dominated by the
pn data, and that the high-energy photon statistics are generally poor
compared to the low-energy statistics, the
inconsistency in the MOS1 photon index has little adverse effect on
the results.  All of the MOS data above 0.7~keV, and pn data above 0.3~keV were utilised during the
spectral fitting, but the residuals from each instrument were examined
separately to judge any inconsistency.

The source spectra were grouped such that each bin contained at least 20
counts. Spectral fitting was performed using {\tt XSPEC v11.2.0} (Arnaud
1996).
Fit parameters are reported in the rest-frame of the object, although most of the
figures remain in the observed-frame.
The quoted errors on the model parameters correspond to a 90\% confidence
level for one interesting parameter (i.e. a $\Delta\chi^2$ = 2.7 criterion).
Luminosities are derived assuming isotropic emission.
The Galactic column density toward PHL~1092 is $N_H$ = (3.6 $\pm$ 0.2)
$\times$
10$^{20}$ cm$^{-2}$ (Murphy et al. 1996).
A value for the Hubble constant of $H_0$=$\rm 70\ km\ s^{-1}\ Mpc^{-1}$ and
a standard cosmology with $\Omega_{M}$ = 0.3 and $\Omega_\Lambda$ = 0.7
has been adopted.  

\subsection{The broad-band spectrum}
\begin{figure}
\psfig{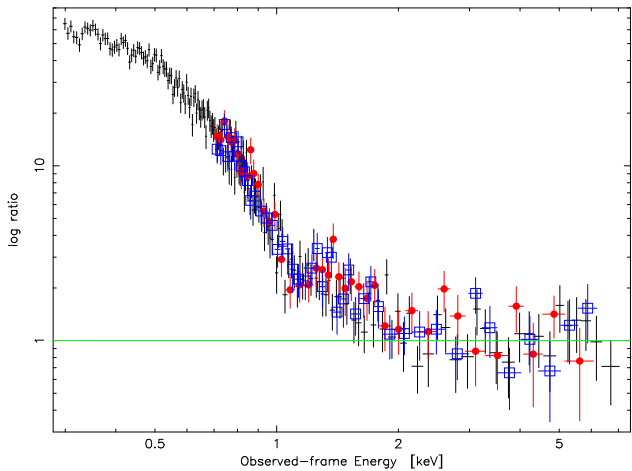}
\caption{The ratio (data/model) resulting from fitting an absorbed power-law
($\Gamma =$ 1.42$^{+0.63}_{-0.50}$) to the 3--7.2~keV EPIC data and extrapolating to 
lower energies.
The black crosses, red dots, and blue squares correspond to the
pn, MOS1, and MOS2 residuals, respectively.  Note that the ratio axis is
logarithmic.
}
\label{pofit}
\end{figure}

A single absorbed power-law is a poor fit to the 0.3--7.2~keV data
($\chi^2$ = 870.5/266 $dof$).  The high statistics below $\sim$ 2~keV
dominate the fit resulting in large residuals at higher energies which
demonstrates the need for multiple continuum components.
For illustrative purposes, a single power-law ($\Gamma =$ 1.42$^{+0.63}_{-0.50}$) modified by Galactic
absorption was fitted to the 3--7.2~keV
EPIC data and extrapolated to lower energies.
The fit
reveals the impressive strength of the soft-excess below $\sim$ 2~keV
(Figure~\ref{pofit}).

Thermal disc models were implemented to fit the soft excess.
Either a single blackbody or a multi-colour
disc blackbody (MCD; Mitsuda et al. 1984; Makishima et al. 1986)
would be a valuable
supplement to the initial power-law fit ($\chi^2$ = 289.6/264 $dof$ and
$\chi^2$ = 290.1/263 $dof$, respectively).
The intrinsic column density was treated as a free parameter and determined
to be insignificant in both models ($<$ 10$^{19}$ cm$^{-2}$)
A double power-law fit and a broken power-law fit were also attempted
to assess whether the soft excess could be attributed to Comptonisation.
Neither the double power-law nor the
broken power-law models were statistically acceptable ($\chi^2$ = 480.0/263
$dof$ and 464.3/263 $dof$, respectively).  In addition, both Comptonisation
models required a high intrinsic column density, on the order of 10$^{21}$
cm$^{-2}$.  Such a large amount of intrinsic cold absorption above the
Galactic value is inconsistent
with the previous findings with $ROSAT$ (Forster \&
Halpern 1996; Lawrence et al. 1997; BBFR).
The high column density derived with the
power-law models is undoubtably manifested to accommodate a soft excess
with intrinsic curvature.

\begin{figure}
\psfig{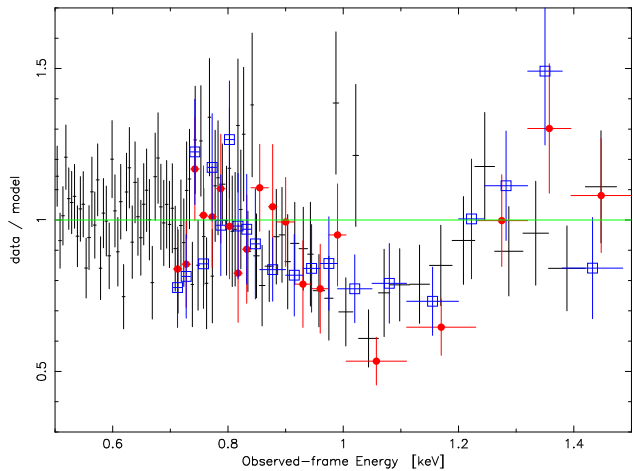}
\caption{An absorption-like feature detected in all three EPIC instruments
at $\sim$ 1~keV ($\sim$ 1.4~keV in the rest-frame).
The continuum is represented
by the best-fit model
described in Sect.~3  and Fig.~\ref{mo}.  See also model (a) in Table~2.
The black crosses, red dots, and blue squares correspond to the
pn, MOS1, and MOS2 residuals, respectively.  The data have been
re-binned for display purposes.
}
\label{abs}
\end{figure}

Regardless of the prefered continuum model, a depression remained in the
residuals at approximately 1~keV (1.4~keV in the rest-frame;
Figure~\ref{abs}).
A Gaussian profile was added to the fits to model a potential absorption line.
The supplementary component was an improvement to both the blackbody plus
power-law fit ($\Delta\chi^2$ = 16.1 for the addition of 3 free
parameters)
and the MCD plus power-law fit ($\Delta\chi^2$ = 31.1 for the addition
of 3 free parameters).
The best-fit energy and equivalent width ($E = 1.43 \pm 0.04$; $EW =
-82^{+19}_{-22}$) are consistent with similar features observed in
other NLS1 (e.g. Leighly 1999b; Vaughan et al. 1999).
The addition of a Gaussian profile was also a significant improvement to
the
Comptonisation models; however, it did not alleviate the requirement
of a high column density in these models.
Replacing the absorption line with an edge also improved the fit, but not as
well as the line model ($\Delta\chi^2$ = 17.5 for the addition of 2 free
parameters to the MCD plus power-law model).  The edge energy is
$E$ $\approx$ 1.33~keV, inconsistent with the strong edges arising
in a warm absorber (e.g. O~{\tt VII} or O~{\tt VIII}).
The residuals
between about 1--1.2~keV
 in Figure~\ref{pofit} show a slight rise.
We examined the possibility that the residuals in the 1--2~keV range could
be due to an emission feature rather than absorption.  Indeed an emission
feature was an improvement to the MCD plus power-law fit ($\Delta\chi^2$ =
25.3 for the addition of 3 free parameters), but not quite as good as an
absorption line.  In addition, the energy and strength of the feature
($E \approx$ 1.97~keV and $EW \approx$ 350~eV) are difficult to reconcile
with the current understanding of warm emission.  Furthermore, residuals
still
remained at $\sim$ 1~keV, and when these residuals were modelled the
parameters of the emission
feature were no longer constrained.

\subsection{Evidence for a reflection component}
\begin{figure}
\psfig{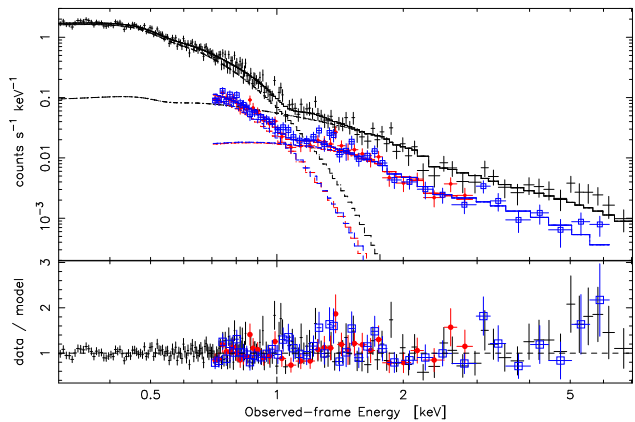}
\caption{The MCD plus power-law and an absorption line model fitted
to the EPIC 0.3--7.2 keV data (see text for details).
In the lower panel the residuals (data/model) are show.
While the fit is statistically acceptable, it is rather poor above
$\sim$2~keV.
The black crosses, red dots, and blue squares correspond to the
pn, MOS1, and MOS2 points, respectively.  For clarity the MOS1 data and
residuals above 3~keV are not included in the figure (see text for detail).
}
\label{dbb}
\end{figure}

In a statistical sense the MCD and power-law model with an absorption line
at
$\sim$ 1.4~keV is a very good fit ($\chi^2$ = 259.0/260 $dof$).
However, when the fit is examined in detail, the residuals indicated
some curvature in the spectra above 2~keV (Figure~\ref{dbb}).

Clearly, the degree of curvature observed at high energies will depend on
how the soft
continuum emission is modelled.  Replacing the low-energy thermal component
with a power-law did not eliminate the high-energy curvature.  In addition,
the quality of the fits were worse ($\chi^2_\nu >$ 1.1), and the models
still required a high level of cold absorption.

Gradual flattening of the power-law toward higher energies has been
observed in some other NLS1, and can be a quality attributed to a dominant
reflection component
(e.g. Fabian \& Vaughan 2003) or partial-covering (Holt et al. 1980).
While partial-covering has been relatively successful in describing
the X-ray spectra of 1H~0707--495 (Boller et al. 2002) and IRAS~13224--3809
(Boller et al. 2003), the situation is more ambiguous in PHL~1092 due to
the more modest statistics and
the absence of a characteristic sharp spectral drop at energies above
7.1~keV (depending on the ionisation state of iron).
We considered an absorption model by fitting the spectrum with a power-law
plus edge.  While the fit to the high-energy spectrum was good,
the overall fit, including the soft X-ray components, was unacceptable
($\chi^2_\nu =$ 1.22).

\begin{figure}
\centerline{
\psfig{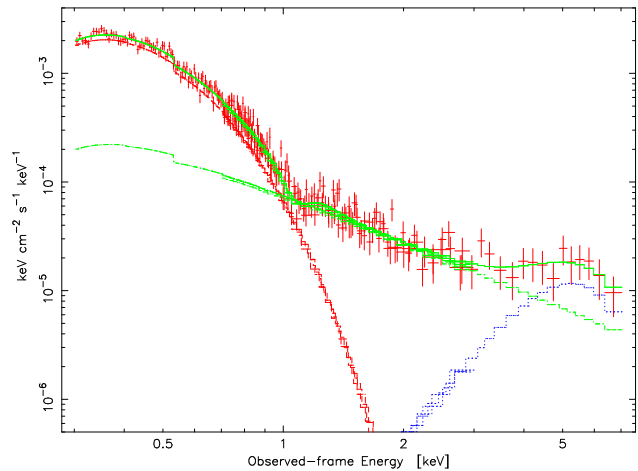}
}
\caption{The unfolded model plotted to the EPIC pn data.  The strong,
broad emission line dominates the high-energy spectrum lending to the
hypothesis that the spectrum is reflection dominated.
}
\label{mo}
\end{figure}

A reflection dominated spectrum has also been suggested for 1H~0707--495
(Fabian et al. 2002) and IRAS~13224--3809 (Boller et al. 2003), as well as
other NLS1 (Ballantyne et al. 2001).  
To test the reflection spectrum hypothesis
a Gaussian profile was added to the fit to emulate an iron emission line.
The improvement to the overall fit was significant ($\Delta\chi^2$ = 14 for
the
addition of 3 free parameters).
The observed spectra can be described by a MCD plus power-law continuum,
with
warm absorption as well as a strong reflection component, all of which is
modified by an amount of neutral absorption
which is consistent with the
Galactic column ($\chi^2$ = 245.0/257 $dof$; Figure~\ref{mo}; model (a) in
Table~1).
The temperature at the inner disc radius is 114 $\pm$ 4 eV.
The power-law has a photon index of 2.55 $\pm$ 0.11.
The absorption line is defined by $E$ = 1.43 $\pm$ 0.04 keV, $\sigma$ =
133$^{+42}_{-37}$, and $EW$ = $-82^{+19}_{-22}$.
The reflection component is modelled by a Gaussian profile with $E$
$\approx$ 6.9 keV,
$\sigma$ $\approx$ 2 keV, and $EW$ $\approx$ 5 keV.
Other models for the reflection component (e.g. {\tt diskline, laor,
pexrav})
were also effective in handling the high-energy curvature.
In these cases the lines were slightly weaker ($EW =$ 2.5--4~keV);
however, the quality
of the data did not justify the use of these more complicated reflection models.
The very strong iron lines suggested for this PHL~1092 observation seem
unphysical, and we will address this issue in Section~5.4.
{\scriptsize
\begin{table*}
\begin{center}
\begin{tabular}{ccccccccccccc}
(1) & (2) & (3) & (4)  & (5) & (6) & (7) & (8) & (9) & (10) & (11) & (12) & (13)  \\
Model & $\chi^2_\nu$ & N$_H$ & $kT$ & $\Gamma_1$ & $\Gamma_2$ & $E_{brk}$ &
$E$ & $\sigma$ & $EW$ & $E$ & $\sigma$ & $EW$ \\
 & (dof) & (10$^{20}$ cm$^{-2}$) & (eV) & & & (keV) & (keV) & (eV) & (eV) &
(keV) & (keV) & (keV) \\
\hline
(a) & 0.95 (257) & $<$0.03 & 114$\pm$4 & 2.55$\pm$0.11 & -- & -- &
1.43$\pm$0.04 & 133$^{+42}_{-37}$ & $-82^{+19}_{-22}$ & 6.9 &
2 & 5 \\

(b) & 0.95 (258) & $<$0.03 & 112$^{+2}_{-5}$ & 2.74$^{+0.15}_{-0.11}$  &
1.49$^{+0.28}_{-0.32}$ & 2.25$^{+0.40}_{-0.31}$ & 1.43$\pm$0.04 &
144$^{+41}_{-37}$ & $-91^{+21}_{-22}$ & -- & -- & -- \\

(c) & 1.07 (257) & 6.6$\pm$0.2 & -- & 4.64$\pm$0.12 & 1.80$\pm$0.26 &
1.88$\pm$0.11 &
1.28$\pm$0.06 & 300$^{+33}_{-28}$ & $-197^{+7}_{-5}$ &
7.2 & 0.71 & 0.52 \\

\hline

\end{tabular}
\caption{Spectral fits to the 0.3--7.2~keV (0.4--10~keV rest-frame) EPIC data.
The two best-fits are models (a) MCD plus power-law continuum plus listed
features, and (b) MCD plus broken power-law and listed features.  Model (c)
is the best-fit Comptonisation continuum model shown for comparison.
Columns (4) to (7) are related to the continuum model: $kT$ is the thermal
temperature at the innermost disc radius; $\Gamma_1$ and $\Gamma_2$ are the photon
indices of the power-law components (two photon indices are required in the case of a broken
power-law); $E_{brk}$ is the energy at which the photon index changes between
$\Gamma_1$ and $\Gamma_2$ in the broken power-law model. Columns (8) to (10)
are the absorption line components: line energy, width, and equivalent width
($E$, $\sigma$, $EW$, respectively), and columns (11) to (13) are the emission
line components (best-fit values): line energy, width, and equivalent width
($E$, $\sigma$, $EW$, respectively).
All models have been modified by line-of-sight Galactic absorption
(3.6 $\times$ 10$^{20}$ cm$^{-2}$).
}
\end{center}
\end{table*}
}

The average 0.3--7.2 keV unabsorbed flux is 1.91 $\times$ 10$^{-12}$ erg
s$^{-1}$ cm$^{-2}$ (1.78 $\times$ 10$^{-12}$ erg s$^{-1}$ cm$^{-2}$ in the
0.3--2~keV band), corresponding to an observed luminosity of
8.8 $\times$ 10$^{44}$ erg s$^{-1}$.
During the 18-day $ROSAT$ HRI monitoring campaign of PHL~1092, BBFR measured
a 0.2--2~keV luminosity between 0.5--7.5 $\times$ 10$^{45}$ erg s$^{-1}$.
Extrapolating our best-fit model from 0.3~keV to 0.2~keV, we estimate an
average 0.2--2~keV luminosity of 1.1$\times$ 10$^{45}$ erg s$^{-1}$.

In comparison with the $ASCA$ luminosities reported by Vaughan et al. (1999;
after correcting for the different cosmology which was assumed), we note
that the intrinsic 0.6--10~keV luminosity is about 12\% higher 
during the {\em XMM-Newton} observation.  However, in the 0.6--2~keV
band the {\em XMM-Newton} observation is $\approx$ 67\% brighter, whereas 
the 2--10~keV luminosity is about 60\% dimmer.  The relative change in
the various X-ray bands suggests long-term spectral variability in PHL~1092; however,
the significance of this result cannot be tested without knowledge of the
flux uncertainties in the $ASCA$ data.

The broad-band continuum could be modelled equally well with
a MCD plus a broken power-law above $\sim$ 2 keV
($\Gamma_1 \approx$ 2.7, $\Gamma_2 \approx$ 1.4, $E \approx$ 3.1~keV),
and a $\sim$ 1.4~keV absorption line ($\chi^2$ = 245.4/258 $dof$).
However, it is difficult to understand the physical significance of the
high-energy break.  Fit parameters for three of the best-fit models are
shown in Table~2 for comparison.  It is apparent from Table~2 that 
the low-energy spectrum is better fit with a blackbody component rather than
a power-law. 

\subsection{The true soft-excess and the need for high-energy curvature}
When fitting the 3--7.2~keV (4.2--10~keV rest-frame) spectrum and extrapolating
downward in energy, as was done for
Figure~\ref{pofit}, there is evidence for an incredibly extreme soft-excess
component, but no support for high-energy curvature.
However, the need for high-energy curvature is seen when the broad-band
spectrum is modelled in full.
The simple power-law plus blackbody fit is significantly improved when
a third component is introduced.
Success is obtained when the simple continuum model is modified with either a
strong, broad $\sim$ 2~keV Gaussian profile, or a high-energy break in the
power-law,
or a strong, broad iron fluorescence line.  
We attempt to demonstrate the curvature in the high-energy spectrum on a
more
basic level.

The 2.5--4~keV rest-frame data were fitted with a single absorbed power-law
($\Gamma \approx$ 2.6).  While the photon index is steep compared to what
is normally measured in AGN it is not atypical of NLS1-type objects (e.g.
Brandt et al. 1997; Porquet et al. 2004).
The 2.5--4~keV region was selected because it was the region
most unlikely to have a significant
contribution from the soft-excess or possible line emission, based on the
general knowledge of AGN X-ray spectra.
\begin{figure}
\psfig{figure=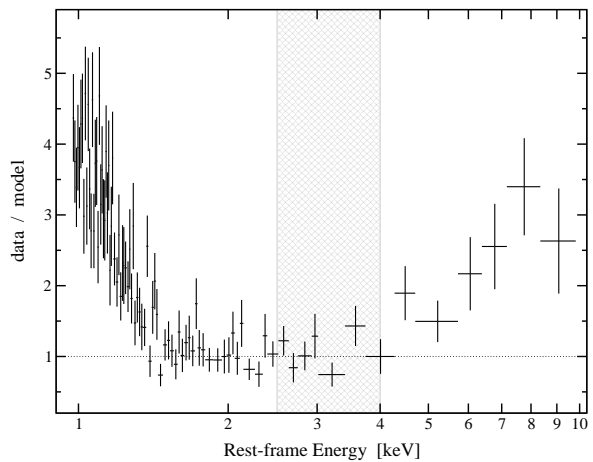,angle=-90,width=8.3cm,clip=}
\caption{The residuals resulting from fitting the rest-frame 2.5--4~keV pn
data (shaded region) with an absorbed power-law, and extrapolating to higher and lower
energies.
The data have been binned up for display purposes only.  Note that the
energy axis is given in the rest-frame.
}
\label{extrap}
\end{figure}
The fit was then extrapolated to lower and higher
energies
as seen in Figure~\ref{extrap} (note the {\em rest-frame} energy axis).  
The extrapolation to
higher energies is clearly poor
indicating that there is a change in the spectral slope somewhere between
2.5--10~keV (rest-frame).
The extrapolation to lower energies is quite reasonable down to
1.5~keV (rest-frame), at which point there is a rather dramatic upturn in
the
residual, likely marking the $true$ onset of the soft-excess.
Pounds \& Reeves (2002) examine the soft-excess in a small sample of
Seyfert~1
galaxies.  One of their conclusions is that the onset of the soft-excess
is related to the 2--10~keV luminosity, such that in more
luminous objects the soft-excess originates at higher energies.  Comparing PHL~1092 to their sample
(correcting for the different cosmology), we determined that a
soft-excess
starting at 1.5~keV (rest-frame) in PHL~1092 is precisely where the onset is expected.

\section{Variability Properties}

\subsection{Extreme X-ray variability}
\begin{figure*}
\centerline{
\psfig{figure=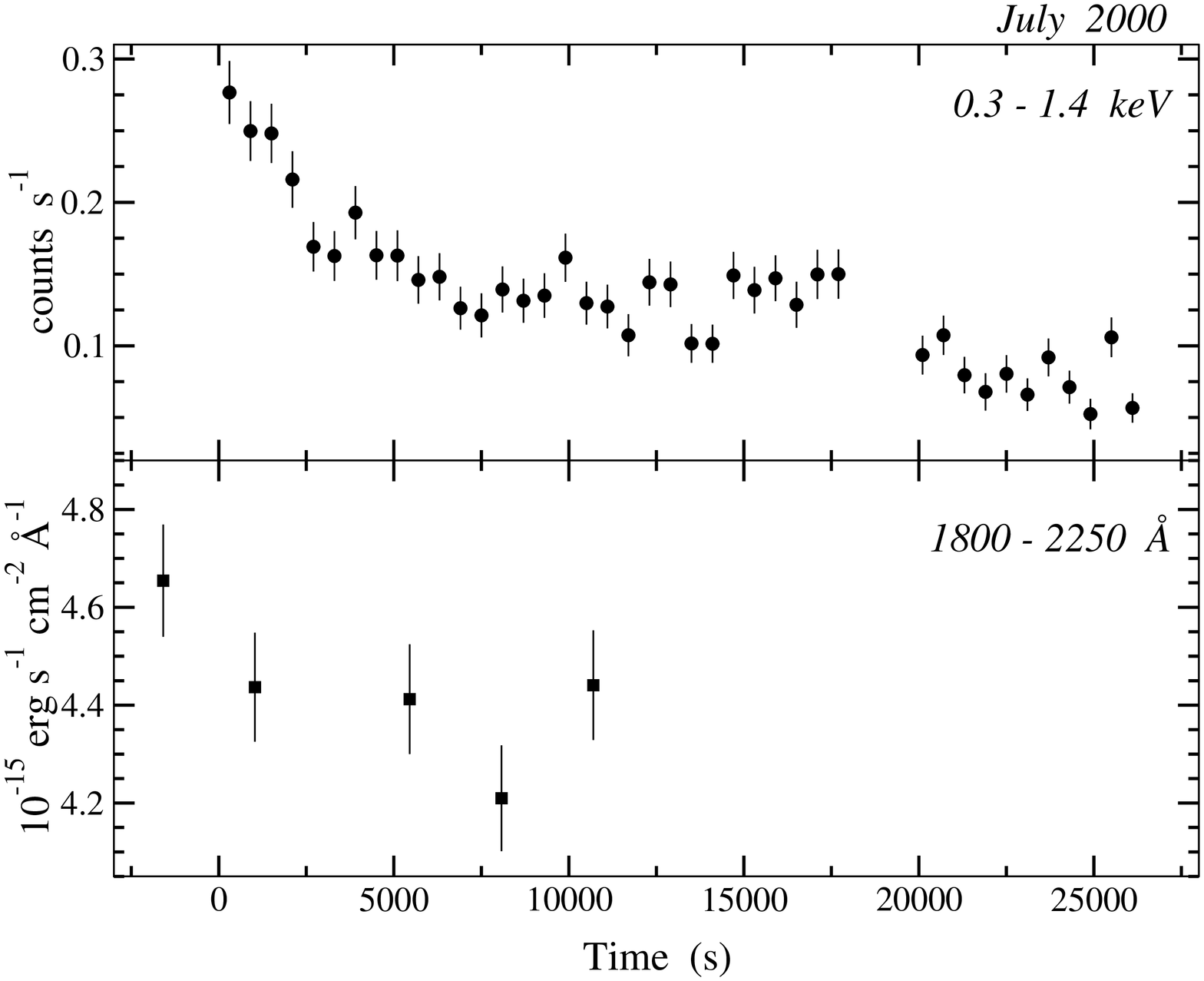,angle=-90,width=9.3cm,clip=}
\nolinebreak
\psfig{figure=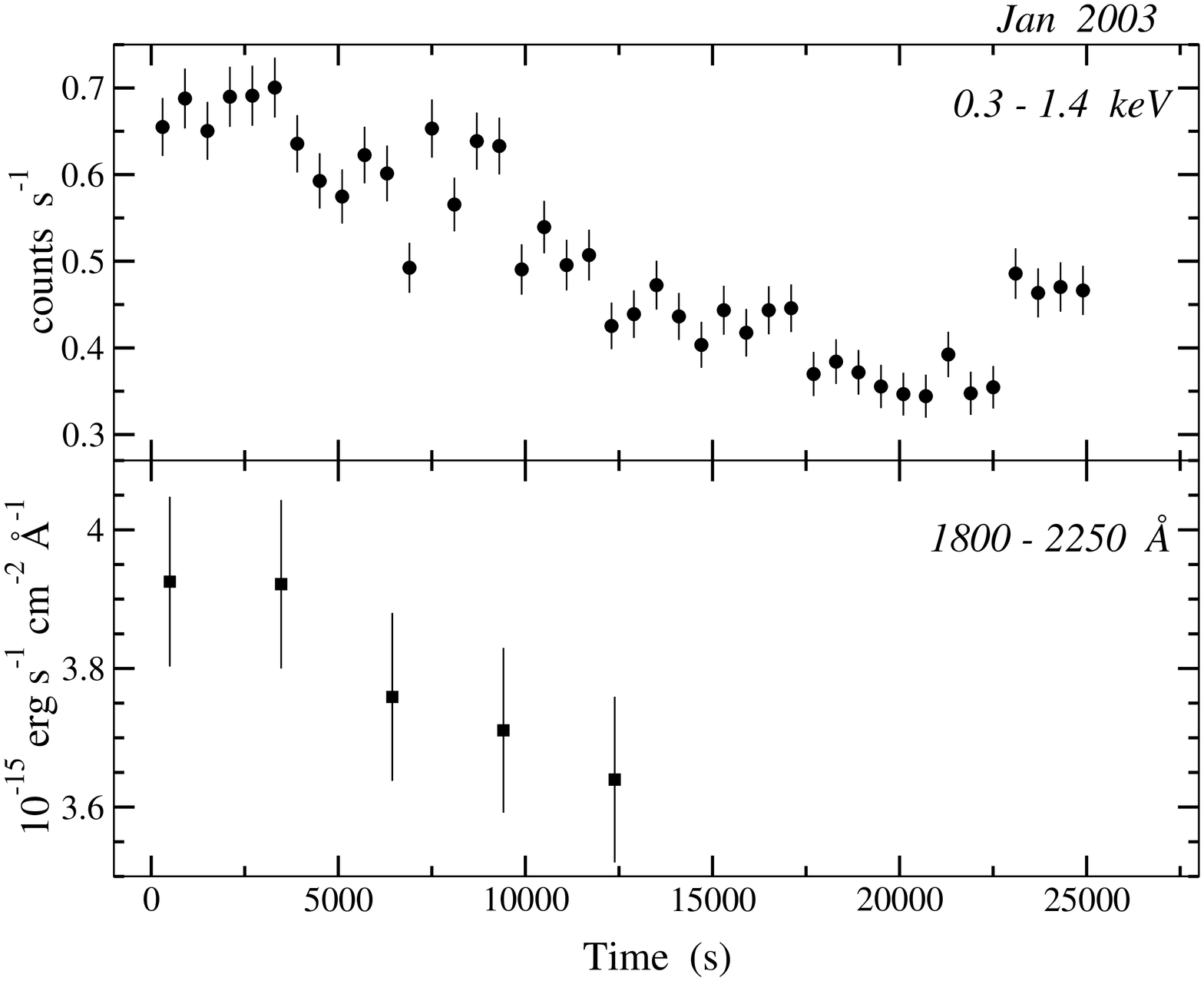,angle=-90,width=9.3cm,clip=}
}
\caption{On the left side are the light curves corresponding to the
July 2000 (GT) observation of PHL~1092.  On the right side are the
January 2003 (AO2) light curves.  In the upper panels on both sides are
displayed the variability in the 0.3--1.4 keV range ($\sim$ 0.4--2 keV in
the rest-frame; 600~s bins).  In the lower panels on both sides are the UVW2 
light
curves
from each observation (1800--2250 \AA~observed-frame; $\sim$ 1290--1600
\AA~rest-frame; 2300~s exposures during GT, and 2640~s exposures during 
AO2).
}
\label{lc}
\end{figure*}

Light curves from the two {\em XMM-Newton} observations of PHL~1092 are
displayed in Figure~\ref{lc}; July 2000 on the left side, and January 2003
on the right (hereafter GT and AO2, respectively).
The 0.3--1.4 keV (0.4--2 keV in the rest-frame)
light curve during the GT observation shows
periods of rapid flux drops and periods of relative quiescence.  Overall,
the X-ray intensity diminishes by nearly 70\% during the 26 ks observation.
During the AO2 observation the average count rate is about twice as high
as during the GT observation.  The 0.3--1.4 keV light
curve\footnote{Including data above 1.4 keV does not contribute
significantly to the total count rate; however background flaring becomes a
factor resulting in gaps in the light curves}
again exhibits a diminishing intensity from the start of the observation
to the end.  However, during the second observation, the flux drop is more
gradual.  The peak intensity falls by about 50\% over the
first 23 ks.

The most rapid event during the AO2 observation occurs during
the final 5 ks.  After reaching a minimum intensity at about 23 ks, the flux
suddenly rises by $\sim$ 30\%, and remains there to the end of the exposure.
Averaging over two low and two high data points in this event (1200~s on 
each
side), the change in
the rest-frame count rate is 0.12 counts s$^{-1}$ in 860 s.  Adopting the 
average
luminosity discussed in the previous section to determine a conversion 
factor between count rate and flux, we find that the event corresponds to a 
luminosity change of $\Delta$L = 2.0 $\times$ 10$^{44}$ erg 
s$^{-1}$.
The luminosity rate of change is $\Delta$L/$\Delta$t $\approx$ 2.3 $\times$
10$^{41}$ erg s$^{-2}$.  Quantifying this rate of luminosity change in terms
of a radiative efficiency, $\eta > 4.8 \times 10^{-43}$$\Delta$L/$\Delta$t, 
(Fabian 1979), we calculate $\eta > 0.11$. 
The measured value of $\eta$ is consistent with 
that expected from a Kerr black hole, but it exceeds the efficiency limit for a
Schwarzschild black hole.
While this is the most rapid event measured during this observation, BBFR found that, along with giant-amplitude flares, smaller amplitude flaring events were also rather common in PHL~1092, during periods of relative low flux (see Figure~1 of BBFR).

\subsection{Simultaneous UV variability}
An interesting discovery from this analysis is the simultaneous UV
fluctuations
of PHL~1092 which may be apparent at both epochs (Figure~\ref{lc} lower panels).
Fitting a constant to both UV light curves results in $\chi^2_{\nu} =$
2.01 and 1.12 for the GT and AO2 observations, respectively; indicating
variability at the $>$ 90\% and $\sim$ 66\% confidence levels.  
While not statistically
significant,
the UV light curves are short ($\sim$ 3 hours), not well-sampled, and of 
modest
signal-to-noise (uncertainties on the level of 5\%).  Despite these
observational constraints, variability trends are seen which appear
consistent with the variability detected in the X-rays (more so during AO2). 
In addition, these short UV light curves are in contrast
to what has been observed in the OM light curves of other NLS1 which normally
show quite constant UV variability curves (e.g.
Gallo et al. 2004a, 2004b).
Unfortunately, relative photometry, to test the reliability of the UV light curves
in Figure~\ref{lc}, is not possible due to the
method in which the OM observation was carried out.  As described by
Mason et al. (2002), in the default OM imaging mode there are five exposures
made up of two image windows each.  While one window remains fixed on the
central X-ray source during all five exposures (PHL~1092 in our case),
the second window is moved
across the CCD in order to maximize the field-of-view coverage.
As a result, there is minimum overlap in the different windows, and no
potential standard star is observed more than once.

Focusing on the AO2 observation, it may appear that the UV and X-rays are
undergoing a gradual decline in flux during the first 13 ks.  During this
time the X-ray count rate drops by about 40\%, whereas the UV flux drops by 
an
average of 7\% with an upper limit of 13\% (not making any corrections for
host galaxy and emission-line contamination).  Without a longer
base line on both
light curves (and a higher signal-to-noise UV light curve) it is impossible to discuss how significant this result is, or
if the two processes are physically connected (e.g. reprocessing, pivoting
power-law, partial-covering).

There also appears to be long-term variability in the 
UV to X-ray spectral slope between the
two epochs.  In a rough approximation, utilising the values from 
the
light curves, we see that the fraction of
the mean X-ray count rate over the mean UV flux ($\div 10^{-15}$) has 
changed
from 0.030 $\pm$ 0.001 during the GT observation, to 0.133 $\pm$ 0.002 during AO2.
We examined whether the change in the UV flux is due to instrumental 
effects
by making note of the UV fluxes of two other objects which were observed at both
epochs (Table~\ref{uvtab}).  
One of these objects is a G5 star (HD~10214); hence its brightness should be
rather stable over the $\sim$ 2 year period.
As we see from Table~\ref{uvtab} the UV flux in these comparison objects does
change, and this could be associated with changes in the intrinsic luminosity
of the sources or instrumental effects.  However, the 
UV flux change in PHL~1092 is larger and opposite to that observed in 
the other two objects; hence the variability appears to be intrinsic to 
PHL~1092.
\begin{table}
\begin{center}
\begin{tabular}{lcccc}
Object & Classification & GT Flux & AO2 Flux  & Change  \\
\hline
HD~10214 & G5 star & 307 $\pm$ 1 & 325 $\pm$ 1 & $+6$\% \\
UGC~00649 & Sab galaxy & 30.7 $\pm$ 0.3 & 33.2 $\pm$ 0.3 & $+8$\% \\
PHL~1092 & AGN & 4.43 $\pm$ 0.05 & 3.79 $\pm$ 0.05 & $-14$\% \\
\hline
\end{tabular}
\caption{Long-term differences in the UV fluxes of identified OM sources.
UVW2 flux densities are given in units of $\times$10$^{-15}$ erg s$^{-1}$ 
cm$^{-2}$ \AA$^{-1}$.
}
\label{uvtab}
\end{center}
\end{table}
A typical $\alpha_{ox}$ derivation is difficult to accomplish without making
some assumptions about the UV spectral slope; hence it will not be calculated
here.

In general, the variability is impressive for an object with an X-ray
luminosity of $\sim$ 10$^{45}$ erg s$^{-1}$.  Moreover, the rapid
variability
seems to extend to lower energies.

\subsection{Spectral variability}

\begin{figure}
\psfig{figure=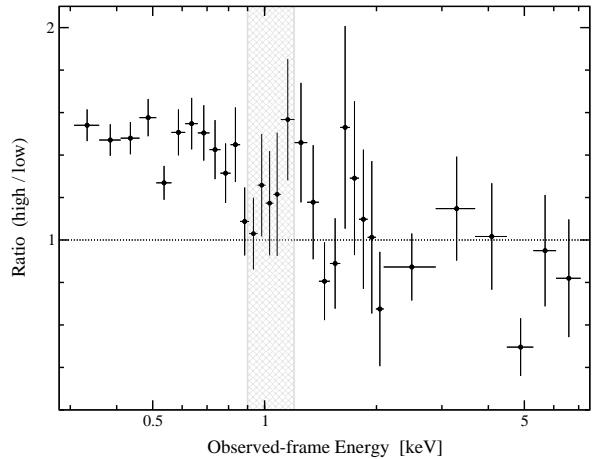,angle=-90,width=8.3cm,clip=}
\caption{The EPIC pn background-subtracted ratio spectrum (high/low).  
To compare the spectrum to a constant a dotted-line is drawn in at
high/low = 1.  While the flux
difference
in the soft emission is remarkable, the variability above 2 keV is
insignificant.  The shaded region marks the approximate position of the
$\sim$ 1.4~keV absorption feature based on Fig.~\ref{abs}.
}
\label{ratio}
\end{figure}

Comparing the $ASCA$ and {\em XMM-Newton} observations there appears to be
long-term X-ray spectral variability.  Considering the OM and EPIC fluxes
from the two {\em XMM-Newton} observations there also appears to be 
significant
variability in the UV/X-ray spectral slope.
However, examining short-term spectral variability within a single
observation is difficult given the degree of high-energy background flaring
during AO2.
We search for flux related spectral variability,  by constructing and 
comparing
a high-flux spectrum
and a low-flux spectrum.
The high-flux state corresponded to all events with
a count rate
greater than 0.5 counts s$^{-1}$ in Figure~\ref{lc} (right panel).  This
was roughly equivalent to selecting all events prior to $\sim$ 12 ks.  The
low-flux state was made up of all remaining events.  In Figure~\ref{ratio}
we have plotted the ratio between the high-flux and low-flux 
background-subtracted spectra.
Note that the ratio spectrum only goes to $\sim$6.5 keV, as the low-flux
spectrum is background dominated above this energy.

For comparison to a constant, a line is drawn at a value of high/low = 1.
The variability is obvious.  At low-energies, $<$ 1 keV (1.4 keV
rest-frame), the variability is remarkable, on the order of 50\%.
However, at energies above $\sim$ 1.4 keV (2 keV in the rest-frame),
Figure~\ref{ratio} shows no indication of significant variability.
Despite the extreme soft X-ray variability seen in Figure~\ref{lc}, it would
seem that the hard X-rays are not particularly variable.

The shaded region in Fig.~\ref{ratio} marks the approximate position of the
$\sim$ 1.4~keV absorption feature based on Fig.~\ref{abs}.
Within the uncertainties, the variability in this region is weak, and most likely related 
to
the underlying continuum rather than the absorption line itself.

\section{Discussion}

\subsection{General findings}
The main results of this study are listed below.
\begin{itemize}
\item[(1)]
The soft excess can be best-fit with a MCD (or blackbody) modified by Galactic
absorption.  An absorption feature was detected at $E \approx$ 1.4~keV
($EW \approx -$82 eV).  The emission gradually flattens
as the energy increases, introducing curvature to the high-energy ($>$
2~keV)
spectra.  This curvature can be modelled empirically with a Gaussian profile
($E \approx$ 6.9~keV, $EW \approx$ 5~keV) or a broken power-law ($\Gamma_1
\approx$ 2.7, $\Gamma_2 \approx$ 1.4, $E \approx$ 3.1~keV in the
rest-frame).  
\item[(2)]
Comparing the high-flux and low-flux spectra suggests significant short-term
spectral
variability.
While the soft component is remarkably variable, the hard component
shows no considerable variability.
Long-term X-ray spectral variability is also suggested by a 
simple comparison between the $ASCA$ and {\em XMM-Newton} luminosities
in different energy bands.  In addition, 
long-term fluctuations in the X-ray/UV spectral slope are indicated 
by comparing the simultaneous 
UV and X-ray fluxes for each {\em XMM-Newton} observation. 

\item[(3)]
The UV and soft X-ray light curves from the two separate {\em XMM-Newton}
observations show
extraordinary short-term variability for such a luminous quasar ($\sim$
10$^{45}$ erg s$^{-1}$).  Radiative efficiency calculations during the
AO2 observation exceed
the limit for accretion onto a Schwarzschild black hole.
\end{itemize}

\subsection{The $\sim$1.4~keV absorption feature}
An absorption line at $\sim$ 1.4~keV has been detected in a number
of NLS1 (e.g. Leighly 1999b; Vaughan et al. 1999; Boller et al. 2003).
Leighly et al. (1997) examined the possibility that such features were
O~{\tt VII}-O~{\tt VIII} edges, significantly blueshifted due to
relativistic outflows.
On the other hand, Nicastro et al. (1999) explain this type of absorption
feature as a blend of resonant absorption lines, mainly due to Fe~L, in a
highly ionized warm absorber.  A strong, steep, soft-excess is a requisite
for the Nicastro et al. interpretation, which appears relevant in PHL~1092.
To the best of our knowledge, PHL~1092 is the most luminous, NLS1 type
object,
in which possible Fe~L absorption has been detected.

The absorption feature in PHL~1092 was not detected during the $ASCA$
observations. 
We simulated $ASCA$-SIS data by using the {\em XMM-Newton} model
with the same number of counts that were obtained during the real $ASCA$ 
observation.  In this case, the absorption line could have been detected,
although it was not clearly seen in the residuals
($\Delta\chi^2 \approx$ 16 for 3 additional free parameters).
This could be indicative of long-term variability in the absorption feature,
but we cannot dismiss the possibility of normalisation and/or profile changes 
in the low-energy continuum which will alter the apparent strength
of any constant line features.
However, since the ionising X-ray luminosity in PHL~1092 is so highly variable it is
possible to consider that the strength of the $\sim$1.4~keV feature is also
time variable.  

As mentioned, during the $ASCA$ observations the total intrinsic 0.6--10~keV luminosity
was about 10\% lower than during the AO2 observation.  
Variability on this scale is typical for PHL~1092 in the course of 
hours.  What may be more relevant here is the strength of the soft-excess
component.  During the $ASCA$ observations the soft-to-total intrinsic luminosity ratio 
($l = 
L_{0.6-2~keV}/L_{0.6-10~keV}$) was $l_{ASCA} \approx 0.56$, where as during AO2
the ratio was $l_{AO2} = 0.84 \pm 0.04$ (uncertainties were not published
for the $ASCA$ observation).
The strength of the soft-excess could be an important consideration if 
the $\sim$ 1.4~keV feature is indeed variable in PHL~1092.

\subsection{The rapid flux variability}
The large-amplitude and rapid X-ray variability is impressive for a quasar
of this luminosity, but more extreme behavior has been observed in
this object before.
It was during the $ROSAT$ HRI monitoring campaign that BBFR critically examined the underlying
assumptions in the derivation of the radiative efficiency limit.  The 
standard
derivation assumes a uniform, spherical emission region with the release
of radiation localized at its centre.  It assumes that relativistic Doppler
boosting and light bending are unimportant. 
Therefore, it is more than likely that the high and sometimes extreme values
of the radiative
efficiency measured in PHL~1092 are due to one or more of these standard 
assumptions being violated.  The spectral properties discussed in Section~3,
in particular the large equivalent width of the iron line,
may suggest that light bending effects are considerable in
this object (see also Section~5.4).

A new development, however, is the apparent simultaneous UV variability, and
{\em lack of} hard X-ray variability.
Without a high quality UV spectrum of PHL~1092, it is difficult to determine
what could be the driving component responsible for the UV variability.  
Assuming that the
UV spectrum of PHL~1092 is similar to that of
I~Zw~1 (Laor et al. 1997), another strong iron emitting NLS1, we can
hypothesis that the source of the variability is continuum since line-emission
is rather modest in this spectral region (1290--1600~\AA~intrinsic).  
Therefore, it is unlikely that the continuum variations are much stronger
than those seen in Figure~\ref{lc} (i.e. the continuum variability is probably 
not suppressed by line emission).
Any models to be considered must, therefore allow for rapid, large-amplitude
X-ray variability and weak UV variability.
Currently, such models cannot be meaningfully discussed without longer base
line and higher quality light curves.

\subsection{Light Bending Model}
A number of empirical models can been used to fit the high-energy curvature
successfully, for example: a reflection-dominated spectrum, 
a high-energy broken power-law, or a broad 2~keV emission line in addition
to the continuum.
The problem with the latter two models is that they lack a clear physical
interpretation.
Recently,
it was demonstrated how a spectrum could be reflection
dominated via light bending effects (Martocchia, Karas, \& Matt 2000;
Fabian \& Vaughan 2003; Miniutti et al. 2003;
Miniutti \& Fabian 2004).  Uttley et al. (2003) consider a strong
reflection component enhanced by light bending effects in describing the
spectrum of the NLS1 NGC~4051 in its extended low-flux state.

The very strong iron line ($EW \approx$ 2.5--5~keV depending on the adopted model) 
suggested for this
PHL~1092 observations seems highly unphysical,
and in contrast to the general belief that the overall strength of the line
diminishes with luminosity (e.g. Reeves \& Turner 2000; but also see
Porquet \& Reeves 2003).
However, the large equivalent width can be understood through light bending. 
Theoretical modelling by Dabrowski \& Lasenby (2001) demonstrate
that in a maximally rotating black hole, with the primary source located
off the rotation axis, and the observer at an inclination of 60\degr,
fluorescence line equivalent widths as high as $\sim$ 5.5~keV could be
realised.
Interestingly, Miniutti \& Fabian (2004) show that the power-law component
could be significantly more variable than the reflection component which
could help explain the apparently constant high-energy X-ray emission in the 
presences
of the highly variably low-energy X-ray emission.
In addition, the high values  
that have been measured for the radiative efficiency in PHL~1092
on numerous occasions can also be understood
if gravitational light bending effects are accounted for.

\subsection{Alternatives to light bending }
We have already discussed that high-energy curvature could be an indication
for partial-covering.  In addition, much of the long-term spectral 
variability
could be explained in terms of partial-covering.  However, a basic
partial-covering model is simply not a good fit to the current spectral data
and the data quality to not warrant the use of more advanced models.

With respect to the amazing variability it is natural to consider 
relativistic
beaming due to jet emission.  PHL~1092 is radio-quiet, and to the best of
our knowledge, only detected in the sensitive NRAO VLA Sky Survey (NVSS;
Condon et al. 1998) at 1.4~GHz at the 1~mJy level.
Furthermore, the spectra at higher energies are not consistent
with jet emission.

There are, of course, other possibilities that could produce some of the
features observed in PHL~1092 (e.g. steep spectrum, extreme $\eta$,
short-lived flares).  For example, an intervening medium, such as: ions
(e.g. Di Matteo et al. 1997), 
magnetic fields (e.g. Merloni \& Fabian 2001), bulk motion of flaring material
(Reynolds \& Fabian 1997; Beloborodov 1999), or a combination 
could explain a number of the effects seen in PHL~1092 over the years.

\section{Conclusions}
PHL~1092 has been long known for its extreme variability and large 
luminosity.
This most detailed X-ray observation to date revealed interesting
spectral features which certainly warrant a deeper look.
The complicated mixture of spectral and timing properties can be explained
$simultaneously$ if we consider light bending effects.
In addition, the
possibly correlated soft X-ray/UV variability demonstrates the potential
for fruitful variability and reverberation mapping studies of PHL~1092.

\section*{Acknowledgements}
Based on observations obtained with {\em XMM-Newton}, an ESA science mission
with 
instruments and contributions directly funded by ESA Member States and
the USA (NASA).
We are very grateful to the people at the
{\em XMM-Newton} Science Operations Centre who took quick action to correct
the problems regarding the GT observation.
Many thanks to the referee, Andrew Lawrence, for a helpful report.
LCG is thankful to Michael Freyberg for help in understanding the 
EPIC calibration problems during the GT observation.
WNB thanks NASA grants NAG5--9924 and NAG5--9933.

\end{document}